# A multi-band atomic candle with microwave-dressed Rydberg atoms

Yafen Cai[1,*], Shuai Shi[1,*], Yijia Zhou[2,*], Jianhao Yu[1], Yali Tian[1], Yitong Li[1], Kuan Zhang[1], Chenhao Du[1], Weibin Li[3,†], and Lin Li[1,‡]

[1]MOE Key Laboratory of Fundamental Physical Quantities Measurement, Hubei Key Laboratory of Gravitation and Quantum Physics, PGMF, Institute for Quantum Science and Engineering, School of Physics, Huazhong University of Science and Technology, Wuhan, 430074, China

[2]Graduate School of China Academy of Engineering Physics, Beijing, 100193, China

[3]School of Physics and Astronomy and Centre for the Mathematics and Theoretical Physics of Quantum Non-equilibrium Systems, University of Nottingham, Nottingham, NG7 2RD, UK

[*]These authors contributed equally: Yafen Cai, Shuai Shi, Yijia Zhou.

[†]weibin.li@nottingham.ac.uk

[‡]li_lin@hust.edu.cn

Stabilizing important physical quantities to atom-based standards lies at the heart of modern atomic, molecular and optical physics, and is widely applied to the field of precision metrology. Of particular importance is the atom-based microwave field amplitude stabilizer, the so-called atomic candle. Previous atomic candles are realized with atoms in their ground state, and hence suffer from the lack of frequency band tunability and small stabilization bandwidth, severely limiting their development and potential applications. To tackle these limitations, we employ microwave-dressed Rydberg atoms to realize a novel atomic candle that features multi-band frequency tunability and large stabilization bandwidth. We demonstrate amplitude stabilization of microwave field from C-band to Ka-band, which could be extended to quasi-DC and terahertz fields by exploring abundant Rydberg levels. Our atomic candle achieves stabilization bandwidth of 100 Hz, outperforming previous ones by more than two orders of magnitude. Our simulation indicates the stabilization bandwidth can be further increased up to 100 kHz. Our work paves a route to develop novel electric field control and applications with a noise-resilient, miniaturized, sensitive and broadband atomic candle.

**Introduction**

The control and detection of microwave (MW) fields are of paramount importance for a wide range of applications, such as radio astronomy [1-3], radar [4, 5], communication [6] and MW quantum technology [7]. The unprecedented level of control over light-atom interactions allows us to nondestructively detect and stabilize amplitudes of free-space MW fields with atomic ensembles. Such an atom-based MW field amplitude stabilizer, termed as an *atomic candle*, is important for a plethora of free-space MW technologies, including material property characterization [8], long-distance environmental monitoring [9], and improving the MW-based quantum operation fidelity [10-12].

In the past two decades, significant progress has been made in realizing atomic candles using atomic ground states [13-16]. The working principle of the ground-state-based atomic candle is the Rabi resonance, in which a phase-modulated resonant MW field couples two hyperfine ground states. The amplitude of the population oscillation is enhanced when the MW Rabi frequency approaches twice the modulation frequency [15, 16]. This enhancement enables effective feedback control for stabilizing the MW field amplitude (proportional to the MW Rabi frequency). However, this protocol only works at MW frequencies corresponding to the ground-state hyperfine splittings of alkali atoms (e.g., 9.2 GHz for Cesium). Moreover, the Rabi-resonance-based protocol suffers from weak MW-atom coupling due to the small magnetic dipole moment, which limits the dynamic range and stabilization bandwidth of the ground-state atomic candle.

Rydberg atoms feature giant dipole moments, long lifetimes [17], and rich level structures, thus holding the promise for a frequency-tunable atomic candle. Atomic candles based on Rydberg atoms have been recognized as a major motivation for the field of Rydberg electrometry

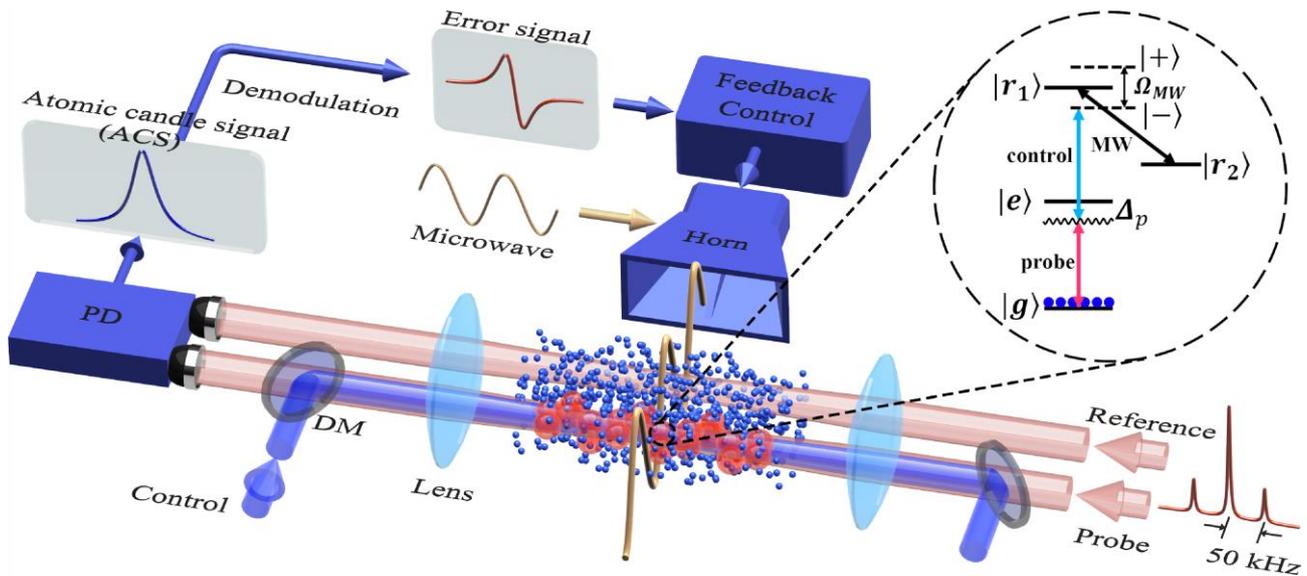

**FIG. 1. Schematic of the Rydberg atomic candle experiment.** The MW-dressed Rydberg spectrum acts as the ACS, which is obtained via a ladder-type EIT scheme. The MW field emitted by a horn resonantly couples Rydberg states $|r_1\rangle$ and $|r_2\rangle$. A balanced detection scheme with a 795 nm reference beam is implemented to further suppress Doppler background and high-frequency intensity noise of the probe field (see **SI** for details). The error signal is generated by introducing a 50 kHz frequency modulation to the probe laser and then demodulating the ACS. The amplitude of the MW field is stabilized by a feedback control loop with the error signal as input. Inset shows the relevant atomic levels of $^{87}$Rb, where $|g\rangle = |5S_{1/2}, F = 2\rangle$, $|e\rangle = |5P_{1/2}, F = 1\rangle$, $|r_1\rangle = |nD_{3/2}\rangle$, and $|r_2\rangle = |(n+1)P_{1/2}\rangle$.

[18-28]. Though promising, experimental realization of a Rydberg atomic candle remains elusive.

In this work, we propose and experimentally implement a new atomic candle scheme using MW-dressed Rydberg atoms. In contrast to the previous realization based on Rabi resonance, our protocol employs the MW-dressed Rydberg spectrum as the atomic candle signal (ACS). This causes large stabilization bandwidth that outperforms its ground-state counterparts by more than two orders of magnitude. Furthermore, our atomic candle is operated at multiple MW frequency bands, which profits from the rich level structures of Rydberg atoms. Large frequency tunability, from MW C-band to Ka-band, is demonstrated in a $^{87}$Rb vapor cell. Our study opens a new window in dressed-state-based atomic candles with excellent prospects for free-space MW applications.

**Protocol**

The prerequisite for MW field amplitude stabilization is to obtain ACS for feedback control, which is an atomic spectrum that is sensitive to the amplitude of MW field. We realize an atomic candle (see Fig. 1) where Rydberg states $|r_1\rangle$ and $|r_2\rangle$ are coupled with a resonant MW field, such that the ACS is obtained from the MW-dressed Rydberg spectrum. As frequency shifts of the MW-dressed Rydberg spectra are proportional to the MW field amplitudes, fluctuations of the MW field amplitudes translate into the frequency shifts. It in turn generates an error signal for the feedback control of the MW field amplitude.

In our experiment, the MW-dressed Rydberg spectrum is obtained through electromagnetically induced transparency (EIT) [29] in a $^{87}$Rb vapor cell, involving atomic ground state $|g\rangle = |5S_{1/2}, F = 2\rangle$, excited state $|e\rangle = |5P_{1/2}, F = 1\rangle$, Rydberg states $|r_1\rangle = |nD_{3/2}\rangle$ and $|r_2\rangle = |(n+1)P_{1/2}\rangle$. The 795 nm probe laser couples $|g\rangle \leftrightarrow |e\rangle$ transition, and the 474 nm control laser is resonant with the $|e\rangle \leftrightarrow |r_1\rangle$ transition. The average Rabi frequencies of probe and control lasers are $\Omega_p = 2\pi \times 7.6$ MHz and $\Omega_c = 2\pi \times 7.3$ MHz. The probe and control lasers are counter-propagating to minimize the Doppler broadening, and their frequencies are stabilized to an ultra-low expansion reference cavity with a finesse of 20 000. Both lasers and MW field are linearly polarized along the same direction. To suppress the DC offset of the EIT spectrum due to Doppler background and high-frequency intensity noise of the probe field, a balanced detection scheme is employed to measure the MW-dressed Rydberg spectrum (see Fig. 1).

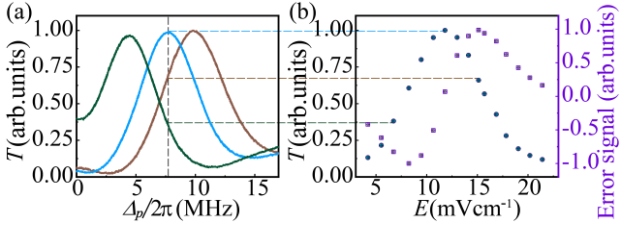

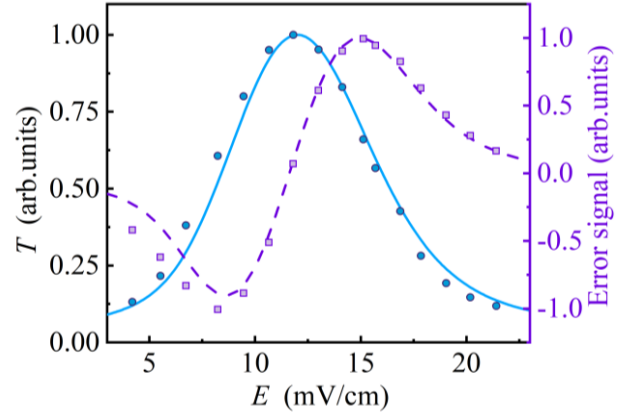

FIG. 2. **Experimental data of the atomic candle signal and error signal.** (a) The spectra of MW-dressed Rydberg states $|+\rangle$ by scanning the probe laser detuning $\Delta_p$. The MW field amplitudes are (green) $6.74 \pm 0.05$ mVcm$^{-1}$, (blue) $12.01 \pm 0.05$ mVcm$^{-1}$, (brown) $15.13 \pm 0.05$ mVcm$^{-1}$, respectively. (b) The atomic candle signal. The target MW field amplitude and corresponding probe laser detuning are $12.01 \pm 0.05$ mVcm$^{-1}$ and $\Delta_p/2\pi = 7.73 \pm 0.03$ MHz, respectively. The detuning is marked by the vertical dashed line in (a). The blue circles show the probe transmission by varying MW field amplitude. The peak transmission is reached at the target MW field amplitude. The corresponding error signal (purple squares) acts as the MW field amplitude discriminator.

FIG. 3. **Comparison between theoretical simulation and experimental data for the atomic candle signal and error signal.** The solid blue curve and blue circles represent the simulated and experimental atomic candle signal with target MW field amplitude of $12.01 \pm 0.05$ mVcm$^{-1}$, which corresponds to a fixed probe laser detuning of $\Delta_p/2\pi = 7.73 \pm 0.03$ MHz. The corresponding simulated and experimental error signals are shown as dashed purple curve and purple squares.

**Atomic candle signal**

As shown in Fig. 1 inset, the MW-dressed Rydberg states $|\pm\rangle = (|r_1\rangle \pm |r_2\rangle)/\sqrt{2}$ lead to transparency windows, with frequencies shifted from the $|g\rangle \leftrightarrow |e\rangle$ transition by $\Delta_\pm = \pm(\lambda_c/\lambda_p)\Omega_{MW}/2$ ($\lambda_p$ and $\lambda_c$ are the wavelengths of the probe and control lasers.), where $\lambda_c/\lambda_p$ is a ratio caused by Doppler averaging [19, 30] and $\Omega_{MW}$ is the Rabi frequency of the MW field. Using the $|+\rangle$ or $|-\rangle$ state, the MW field amplitude $E$ is directly mapped into the MW-dressed Rydberg spectrum through $E = \hbar\Omega_{MW}/\mu_{r_1r_2} = 2(\lambda_p/\lambda_c)\hbar|\Delta_\pm|/\mu_{r_1r_2}$, where $\hbar$ is the reduced Planck constant and $\mu_{r_1r_2}$ is the transition dipole moment between the Rydberg states. Figure 2(a) displays the transmission spectra of the $|+\rangle$ states as a function of $\Delta_p$, where $\Delta_p$ is the probe laser detuning. It illustrates that the frequency shift of the MW-dressed Rydberg spectrum is proportional to the MW field amplitude. The linewidths of the transmission spectra mainly result from Doppler and transit time broadening [31].

The response of the ACS to the MW field amplitude is shown in Fig. 2(b). The peak of the ACS signifies the target MW field amplitude, which is proportional to the probe laser detuning $\Delta_p$. Detuning $\Delta_p/2\pi = 7.73 \pm 0.03$ MHz [vertical dashed line in Fig. 2(a)] corresponds to the target MW field amplitude at $12.01 \pm 0.05$ mVcm$^{-1}$ [ACS peak in Fig.2(b)]. As illustrated by the data connected by the dashed horizontal lines between Fig. 2(a) and Fig. 2(b), the ACS reaches the maximum at the target MW field amplitude and decreases away from the target value. The data indicate that the Rydberg atomic candle can be realized by stabilizing the MW field amplitude to the peak of the ACS.

In order to generate the error signal, a 50 kHz frequency modulation is added to the probe laser. We emphasize that the modulation bandwidth of the Rydberg atomic candle can, in principle, be improved up to 4 MHz [see Supplementary Information (**SI**) for details]. Different from the previous experiment where the MW field amplitude is modulated, our approach can avoid disturbing the MW field. The detected ACS is then demodulated to extract an effective error signal that is sensitive to the amplitude (see **SI** for details). The error signal [purple squares in Fig. 2(b)] is applied to stabilize the MW field amplitude to the ACS peak by compensating the external fluctuation.

**Modeling**

To understand the ACS and error signal, we model dynamics of system with the quantum master equation ($\hbar \equiv 1$),

$$\dot{\rho} = \mathcal{L}[\rho] = -i[H, \rho] + \mathcal{D}[\rho], \quad (1)$$

where $\rho$ is the density matrix of the atom, and $H = -\tilde{\Delta}_p(t)(\sigma_{ee} + \sigma_{r_1 r_1} + \sigma_{r_2 r_2}) + \Omega_p/2(\sigma_{ge} + \sigma_{eg}) + \Omega_c/2(\sigma_{er_1} + \sigma_{r_1 e}) + \Omega_{MW}/2(\sigma_{r_1 r_2} + \sigma_{r_2 r_1})$ is the Hamiltonian with $\tilde{\Delta}_p(t) = \Delta_p + \delta_m \sin(\omega_m t)$, $\Omega_p$ and $\Omega_c$ being the time-dependent detuning, Rabi frequencies of the probe and control lasers, respectively. The atomic projection operators are $\sigma_{ij} = |i\rangle\langle j|$, where $|i\rangle$ and $|j\rangle$ referring to states $|g\rangle, |e\rangle, |r_1\rangle$ or $|r_2\rangle$. The dissipative operator, $\mathcal{D}[\rho]$, consists of spontaneous decay and dephasing [see Supplementary Information (**SI**) for details]. The detuning is modulated at a low frequency $\omega_m \ll \gamma_e$ (in this work, $\omega_m/2\pi = 50$ kHz, and $\gamma_e/2\pi = 5.7$ MHz, where $\gamma_e$ is the natural linewidth of state $|e\rangle$.) and with a small amplitude $\delta_m$, such that the density matrix of the system (including the population and coherence terms) oscillates around a steady value. For single frequency modulation, the density matrix $\rho(t)$ is expanded to be $\rho(t) = \sum_{n=-\infty}^{+\infty} \rho^{(n)} \exp[in\omega_m t]$ where $\rho^{(n)}$ is the time-independent, n-th order Fourier component of the density matrix. Substituting the expansion to the master equation, solutions to $\rho^{(n)}$ can be obtained order by order through iteration (see **SI** for details).

Assuming the atomic density $\mathcal{N}$ is homogeneous in the vapor cell, the transmission of the probe laser is given by Beer-Lambert law [32],

$$T = \exp\left[-\frac{4\pi \mathcal{N} L |\mu_{ge}|^2}{\hbar \epsilon_0 \lambda_p \Omega_p} \operatorname{Im} \rho_{ge}(t)\right], \quad (2)$$

where $L$ is the length of the vapor cell, and $\mu_{ge}$ is the dipole moment of the transition $|g\rangle \leftrightarrow |e\rangle$, $\epsilon_0$ is the

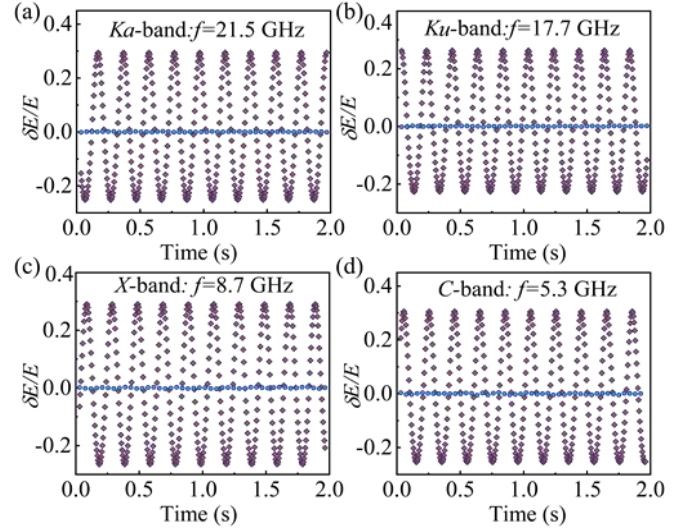

**FIG. 4. Performance of the Rydberg atomic candle.** Frequency tunability of the Rydberg atomic candle from MW C-band to Ka-band is shown (band is labelled in the figure). The corresponding transitions are $|47D_{3/2}\rangle \leftrightarrow |48P_{1/2}\rangle$, $|50D_{3/2}\rangle \leftrightarrow |51P_{1/2}\rangle$, $|63D_{3/2}\rangle \leftrightarrow |64P_{1/2}\rangle$, $|74D_{3/2}\rangle \leftrightarrow |75P_{1/2}\rangle$, respectively. The resilience to disturbance is demonstrated by applying a 5 Hz disturbance in MW field amplitude. The relative amplitude of the MW field as a function of time is shown in each panel, when the feedback control loop is open (brown diamonds) and closed (blue circles).

vacuum permittivity, and $\rho_{ge}(t)$ is the quasi-stationary solution to Eq. (1). The ACS is the stationary part $\bar{T}$ of Eq. (2), which is obtained by temporally averaging over a few modulation periods in the experiment. Taking the Maxwell-Boltzmann distribution at 60 °C and the experimental condition into account, the ACS can be calculated numerically, shown as the solid blue curve in Fig. 3, which agrees with the experimental data in Fig. 2(b).

The error signal is proportional to the coefficient of the first-order oscillatory term of Eq. (2),

$$\text{Error signal} \propto \bar{T} \times \frac{B}{A} I_1 \left(-\frac{4\pi \mathcal{N} L |\mu_{ge}|^2}{\hbar \epsilon_0 \lambda_p \Omega_p} A\right), \quad (3)$$

where $I_1(x)$ is the modified Bessel function of the first kind, $A = \sqrt{|\rho_{ge}^{(1)}|^2 + |\rho_{ge}^{(-1)}|^2 - 2\operatorname{Re}[\rho_{ge}^{(1)} \rho_{ge}^{(-1)}]}$, and $B = \operatorname{Re}[\rho_{ge}^{(1)} - \rho_{ge}^{(-1)}]$. As shown in Fig. 3 (dashed purple curve) the error signal is consistent with the experiment

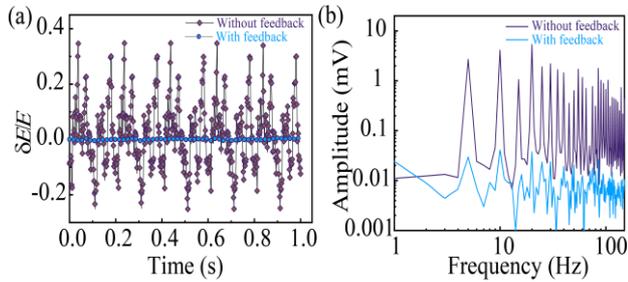

**FIG. 5. Noise-resilience of the Rydberg atomic candle.** (a) The noise resilience is verified by superimposing a disturbance at random frequency spanning from 5 Hz to 150 Hz to the MW field amplitude. (b) The Fourier spectrum of the noise disturbance with feedback loop open and closed. For the disturbance below 100 Hz, the Rydberg atomic candle realized more than two orders of magnitude suppression in the MW field amplitude fluctuation.

data and approximately the derivative of the ACS near the transparent window, which ensures that the error signal is linearly dependent on the deviation from the target MW field amplitude within $\pm 2.3$ mV/cm, and guarantees an efficient feedback control of the MW field amplitude.

**Performance of the Rydberg atomic candle**

A remarkable feature of the Rydberg atomic candle is its excellent frequency tunability. In the following, we will demonstrate that MW fields from C-band to Ka-band can be stabilized. The MW C (X, Ku, Ka)-band covers the frequency range of 3.9-6.2 GHz (6.2-10.9 GHz, 10.9-20 GHz, 20-36 GHz). We employ $|nD_{3/2}\rangle \leftrightarrow |(n+1)P_{1/2}\rangle$ transition with $n = 47, 50, 63,$ and $74$, whose operation frequencies are 21.5 GHz in Ka-band, 17.7 GHz in Ku-band, 8.7 GHz in X-band and 5.3 GHz in C-band. As the level spacing of the neighboring Rydberg states scales as $\sim n^{-3}$, one expects to achieve S- and lower bands with Rydberg states $(n \geq 80)$, or terahertz with Rydberg states $(n \leq 30)$(ref.[33]). Moreover, the atomic candle can even be extended to quasi-DC electric fields by exploring the large static polarizability of the Rydberg states [25, 28].

To quantitatively evaluate the resilience to disturbance of the Rydberg atomic candle, a 5 Hz sinusoidal variation is superimposed on the MW field amplitude. The relative MW field amplitude variations $(\delta E/E)$ are shown in Fig. 4, which are significantly suppressed by closing the feedback control loop. To quantify the MW field amplitude fluctuation, an empirical relative deviation $\Sigma$ is adopted, which satisfies $|\delta E/E| < \Sigma$ for the 90% of the time. As shown in Fig. 4, for Ka (Ku, X, C)-band MW at 21.5 (17.7, 8.7, 5.3) GHz, the deviation $\Sigma$ of the sinusoidal fluctuation without feedback from the ACS [Fig. 4(b) (brown diamonds)] is 27.7% (24.8%, 27.7%, 28.7%). In comparison, the deviation is suppressed to $\Sigma' = 0.19\%$ (0.20%, 0.26%, 0.51%) with the feedback control loop closed (blue circles).

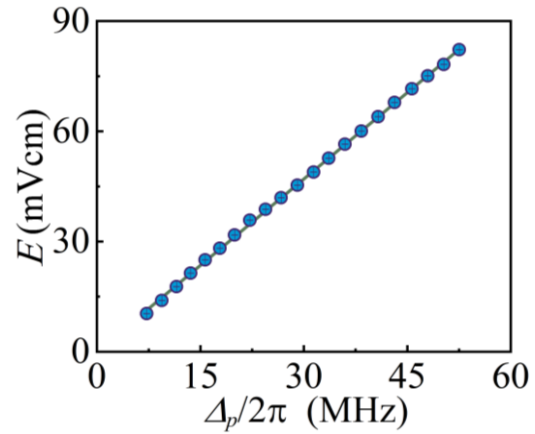

**FIG. 6. Dynamic range of the Rydberg atomic candle at 14.8 GHz.** MW field amplitudes, ranging from 11.3 mVcm$^{-1}$ to 82.3 mVcm$^{-1}$, can be stabilized by changing the probe laser detuning.

The *efficacy* $\varepsilon = \Sigma/\Sigma'$ quantifies the suppression in the MW field amplitude deviation after the feedback is applied [13]. The efficacies associated with the examples in Fig. 4 are $146, 124, 105$ and $56$ for Ka-, Ku-, X-, and C-bands, respectively. The slightly lower efficacy in C-band results from the reduced signal to noise ratio (SNR) of the ACS in higher-lying Rydberg states, which is mainly due to the increasingly stronger interactions when higher Rydberg states are employed. When the atomic blockade radius becomes considerable to the atomic spacing, less atoms are likely to satisfy the EIT condition for the generation of error signal. Technically, optimization of the atomic density, the probe and control lasers can mitigate this interaction induced effect and increase SNR [34]. However, this effect is not significant for the carrier frequencies above C-band.

Although similar stabilization efficacy has been demonstrated with the ground-state atomic candle, the bandwidth is only 0.1 Hz due to its low ACS modulation frequency [9]. The ground-state atomic candle employs atoms' second-harmonic response to the phase-modulated

MW field as its ACS signal. Hence the modulation frequency is ultimately limited by the small ground-state Rabi frequency.

The Rydberg atomic candle overcomes this limitation as the ACS is encoded in the MW-dressed Rydberg spectrum, whose modulation frequency is moreover not bound by the MW Rabi frequency. From MW C- to Ka-bands, stabilization bandwidth up to 5 Hz is achieved (see Fig. 4). To verify the noise resilience of the Rydberg atomic candle, sinusoidal fluctuations at random frequency spanning from 5 Hz to 150 Hz are superimposed on MW field amplitude, and the stabilization effects of the Rydberg atomic candle in time and frequency domain are shown in Fig. 5. Our experiments achieve more than two orders of magnitude suppression of the MW field amplitude variations at frequencies below 100 Hz. Therefore, the stabilization bandwidth of the Rydberg atomic candle is more than 100 Hz.

We then theoretically evaluate the limit of the stabilization bandwidth of the ACS. According to our simulation, the derivative of the error signal against the MW field amplitude decreases when the modulation frequency increases. Typically, a 3 dB deduction signifies the modulation bandwidth being approximately 4 MHz (see **SI** for details). In principle, the stabilization bandwidth can be further increased to $\sim 100$ kHz, which is only limited by the atomic relaxation time [35-37]. Larger stabilization bandwidth allows Rydberg atomic candle to suppress higher frequency noises, which is critical for practical applications.

The MW field amplitude can be stabilized at the desired value by choosing proper probe laser detuning to match the frequency shift of the target MW field amplitude. Figure 6 shows a typical dynamic range at 14.8 GHz using $|53D_{3/2}\rangle \leftrightarrow |54P_{1/2}\rangle$ transition, where the stabilized MW field amplitude depends linearly on the probe laser detuning with a coefficient of $(2\lambda_p/\lambda_c)h/\mu_{r_1 r_2} = 1.55$ mV/cm/MHz. The deviation of the stabilized MW field amplitude depends on both lasers frequency deviations and MW detuning. The resulting uncertainty of the stabilized MW field amplitude is about 0.05 mVcm$^{-1}$ (see SI for a detailed explanation). Due to the giant dipole moment, the Rydberg atomic candle is more sensitive to weak MW field than the ground-state one. The ability to stabilize weak MW fields is limited by the $\sim 6$ MHz linewidth of the ACS to $\sim 11$ mVcm$^{-1}$. The linewidth of the ACS depends on the residual Doppler broadening [38] and power broadening in current two-photon excitation scheme. ACS with narrower spectrum linewidth can be obtained by using cold atoms or Doppler-free excitation protocol [39].

**Conclusion and outlook**

We have proposed and experimentally implemented a novel atomic candle for stabilizing the amplitude of free-space MW fields. Our Rydberg dressed-state atomic candle protocol is conceptually new and exhibits unique advantages. First, existing Rabi-resonance-based schemes are limited by stabilization bandwidth, as the amplitude of atoms' second harmonic response to the MW phase modulation is used as ACS. In the present work, Rydberg dressed-state spectrum serves as ACS directly, which can improve the stabilization bandwidth by orders of magnitude. Second, previous schemes only work at MW frequencies corresponding to the ground-state hyperfine splittings of alkali atoms. Our scheme features multi-band MW field stabilization that is enabled by abundant Rydberg level structures. Third, error signals of previous schemes are generated through modulating and then disturbing the MW field amplitude, while the dressed-state protocol can modulate the Rydberg excitation laser frequency instead, and hence avoid disturbing the MW field. Furthermore, the dressed-state atomic candle scheme is not limited to Rydberg transitions and can be used for MW ground-states transitions and optical transitions.

The realization of a novel Rydberg atomic candle opens up new perspectives for important free-space MW applications. For example, it allows the precise measurement of material properties such as absorption and refractive index. Furthermore, MW-based long-distance environmental monitoring can be achieved by using Rydberg atomic candle to stabilize the remote and local MW field amplitude to atomic standard. What's more, MW with good stability can find important applications in improving the plasma emission spectra precision [41]. In addition, MW-based quantum operation fidelity can be greatly improved by using the atoms themselves to suppress the MW field amplitude fluctuation.

**Acknowledgements**

This work was supported by the National Key Research and Development Program of China under Grants No. 2021YFA1402003, and by the National Natural Science Foundation of China (Grant No. 12004127, No. 12004126, No. 12104173 and No. 12005067). W. L. acknowledges support from the EPSRC through Grant No. EP/R04340X/1 via the QuantERA project "ERyQSenS", the UKIERI-UGC Thematic Partnership (IND/CONT/G/16-17/73), and the Royal Society through the International Exchanges Cost Share award No. IEC\NSFC\181078. Y. Z. is supported by National *Natural Science Foundation of China (Grant No.* 12088101), and NSAF (Grant No. U1930403).